Quantum siphoning of finely spaced interlayer excitons in reconstructed $MoSe_2$/$WSe_2$ heterostructures

Mainak Mondal[1], Kenji Watanabe[2], Takashi Taniguchi[3], Gaurav Chaudhary[4], Akshay Singh[1*]

[1]Department of Physics, Indian Institute of Science, Bengaluru, India

[2]Research Center for Functional Materials, National Institute for Materials Science, Japan

[3]International Center for Materials Nanoarchitectonics, National Institute for Materials Science, Japan

[4]TCM Group, Cavendish Laboratory, University of Cambridge, Cambridge, United Kingdom

Corresponding Author: *aksy@iisc.ac.in

**Abstract**

Atomic reconstruction in twisted transition metal dichalcogenide heterostructures leads to mesoscopic domains with uniform atomic registry, profoundly altering the local potential landscape. While interlayer excitons in these domains exhibit strong many-body interactions, extent and impact of quantum confinement on their dynamics remains unclear. Here, we reveal that quantum confinement persists in these flat, reconstructed regions. Time-resolved photoluminescence spectroscopy uncovers multiple, finely-spaced interlayer exciton states (~ 1 meV separation), and correlated emission lifetimes spanning sub-nanosecond to over 100 nanoseconds across a 10 meV energy window. Cascade-like transitions confirm that these states originate from a single potential well, further supported by calculations. Remarkably, at high excitation rates, we observe transient suppression of emission followed by gradual recovery, a process we term "quantum siphoning". Our results demonstrate that quantum confinement and competing nonlinear dynamics persist beyond the ideal moiré paradigm, potentially enabling applications in quantum sensing and modifying exciton dynamics via strain engineering.

**Introduction**

Twisted two-dimensional (2D) transition metal dichalcogenide heterostructures (TMD HS) have emerged as a versatile platform for exploring strongly correlated phenomena, including Hubbard model physics, Wigner crystallization, and excitonic superfluidity, owing to their tuneable moiré superlattices and pronounced light–matter interactions[1–3]. Most interpretations to-date rely on a rigid moiré lattice model, wherein the periodic potential landscape dictates optical and electronic properties. However, when the twist angle and lattice mismatch between constituent layers are small (e.g. $MoSe_2/WSe_2$), lattice reconstruction becomes energetically favourable, leading to domains with energetically preferred atomic arrangements[4–6]. The resulting landscape is further modulated by defect pinning and inhomogeneous strain, leading to complex structural patterns such as 2D domains, 1D stripe-like features, and zero-dimensional confinement pockets[7–9]. These reconstructed domains fundamentally deviate from ideal moiré potentials, raising new questions about exciton localization and dynamics in these systems.

Photoluminescence (PL) from reconstructed domains often shows a single, narrow interlayer exciton (IELX) emission peak, implying the absence of resolvable quantized states[8,10]. This is in stark contrast to ideal moiré systems (~ 2°-twisted $MoSe_2/WSe_2$ HS), where IELX exhibit discrete energy levels spaced by approximately 20 meV, consistent with strong quantum confinement[11]. This discrepancy raises the fundamental question of the impact of quantum confinement in reconstructed systems. Theoretical studies suggest that such domains may host laterally extended excitonic states and ultra-flat energy bands arising from domain-scale electronic hybridization and relaxation[12,13]. Moreover, strong exciton–phonon coupling in reconstructed regions may underlie complex phenomena such as negative diffusion, long-range coherence, and the emergence of many-body excitonic phases[14–16]. Despite these intriguing findings, direct experimental signatures of quantum confinement and modification of excitonic dynamics in reconstructed domains have remained elusive.

Here, we uncover multiple finely-spaced IELX energy levels in $MoSe_2/WSe_2$ HS, under low-repetition-rate excitation ($\leq$ 8 MHz). On the other hand, continuous-wave and high-repetition-rate excitation (80 MHz) yields a single emission peak, consistent with prior reports for PL from reconstructed 2D domains. Time-resolved PL (TRPL) measurements reveal dramatic variation in decay lifetimes of these levels, from sub-nanosecond to over 100 nanoseconds, within a narrow energy window (~ 10 meV). We also observe cascade-like emission features indicative of sequential filling of quantized excitonic states. Calculations of quantized energy levels in large mesoscopic domains agree with the observed ~ 1 meV energy-level spacing. These observations demonstrate the persistence and impact of quantum confinement in mesoscopic reconstructed domains. Furthermore, by varying excitation power and repetition rate, we identify a pronounced transient suppression in PL intensity, a phenomenon we term "quantum siphoning effect". We attribute this nonlinear effect to excitonic scattering from bright K-valley states into non-radiative Q and Γ valleys, a concept of broad relevance in the 2D field. Quantum siphoning provides a way towards novel excitonic responses, for example, via strain-engineering to control valley scattering pathways. The lifetime variation over two orders of magnitude potentially resolves the longstanding discrepancies in reported excitonic dynamics in these systems, and can be exploited for quantum sensing applications. Our results reveal a rich landscape of radiative and non-radiative pathways, offering new insight into quantum many-body dynamics in structurally reconstructed 2D materials.

**Signature of finely spaced multiple energy levels in mesoscopic reconstructed domains**

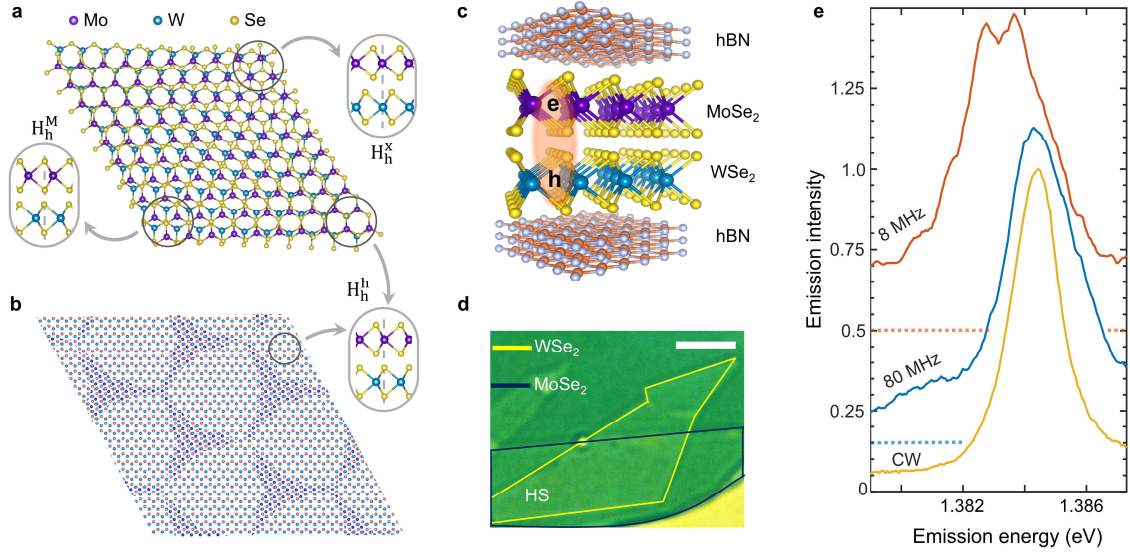

**Fig. 1 | Repetition rate dependent emission from reconstructed domains.** (a) Schematic representation of different atomic high-symmetry zones in H-type $MoSe_2$/$WSe_2$ moiré system. (b) Illustration of atomic reconstruction showing increase of $H_h^h$ registry size. (c) Schematic of $MoSe_2$ and $WSe_2$ monolayer hosting electron and hole, respectively, to form an interlayer exciton with hBN encapsulation. (d) Optical image of the sample. Scale bar is 3 μm. (e) Interlayer exciton emission (IELX) for continuous wave (CW) laser excitation, and 80 and 8 MHz repetition-rate laser excitation, as labelled. Data is vertically shifted for better visualization. Horizontal red and blue dotted line indicate the offset level of 8 and 80 MHz data, respectively.

We prepare the HS by stacking exfoliated $MoSe_2$ and $WSe_2$ monolayers with hBN encapsulation on a $SiO_2$/Si substrate using the dry transfer technique followed by annealing at 250 ºC in an inert nitrogen atmosphere (Fig. 1d, see methods for more details)[17]. Room-temperature PL mapping revealed quenched intralayer emission from the HS region, confirming efficient charge transfer between layers (Fig. 1c), and excellent interface quality (Supplementary Section 1)[18]. Using second harmonic generation, we verified the H–type stacking (Supplementary Section 2)[19,20]. In H-type $MoSe_2$/$WSe_2$ system, the three atomic registries are $H_h^h$ (hexagon on hexagon), $H_h^x$ (chalcogen atom on hexagon) and $H_h^M$ (metal atom on hexagon), as shown in Fig. 1a. Stacking energies follow a hierarchical order: $H_h^h < H_h^x < H_h^M$, resulting in large $H_h^h$ domains formation following reconstruction (Fig. 1b)[21,22]. Next, we measure PL at low temperature ~ 4 K using a continuous wave (CW) laser and a pulsed laser (with 80 MHz repetition rate), shown in Fig. 1e. PL in both cases shows similar IELX characteristics such as peak linewidth, energy position, temperature and power-dependent nature (Supplementary Section 3, 4 and 5), as previously observed from large area 2D $H_h^h$ domains in reconstructed H-type MoSe$_2$/WSe$_2$ HS[9,14,16,23,24]. On the other hand, we observe significantly different IELX PL for 8 MHz excitation (with matched pulse peak power of 80 MHz excitation, and 5 µW of average power), compared to CW and 80 MHz excitation (Fig. 1e). The PL evolves into a multi-peak emission feature for excitation with ≤ 8 MHz repetition rate (Supplementary Section 7). We also observe an overall broadening of the emission and an increase in emission intensity in the lower-energy emission region, compared to the 80 MHz excitation. This multi-peak structure indicates finely spaced IELX energy levels and complex emission dynamics, challenging the previous understanding of a single emission peak from the large reconstructed domains[8].

**Dynamical corelation between finely spaced energy levels**

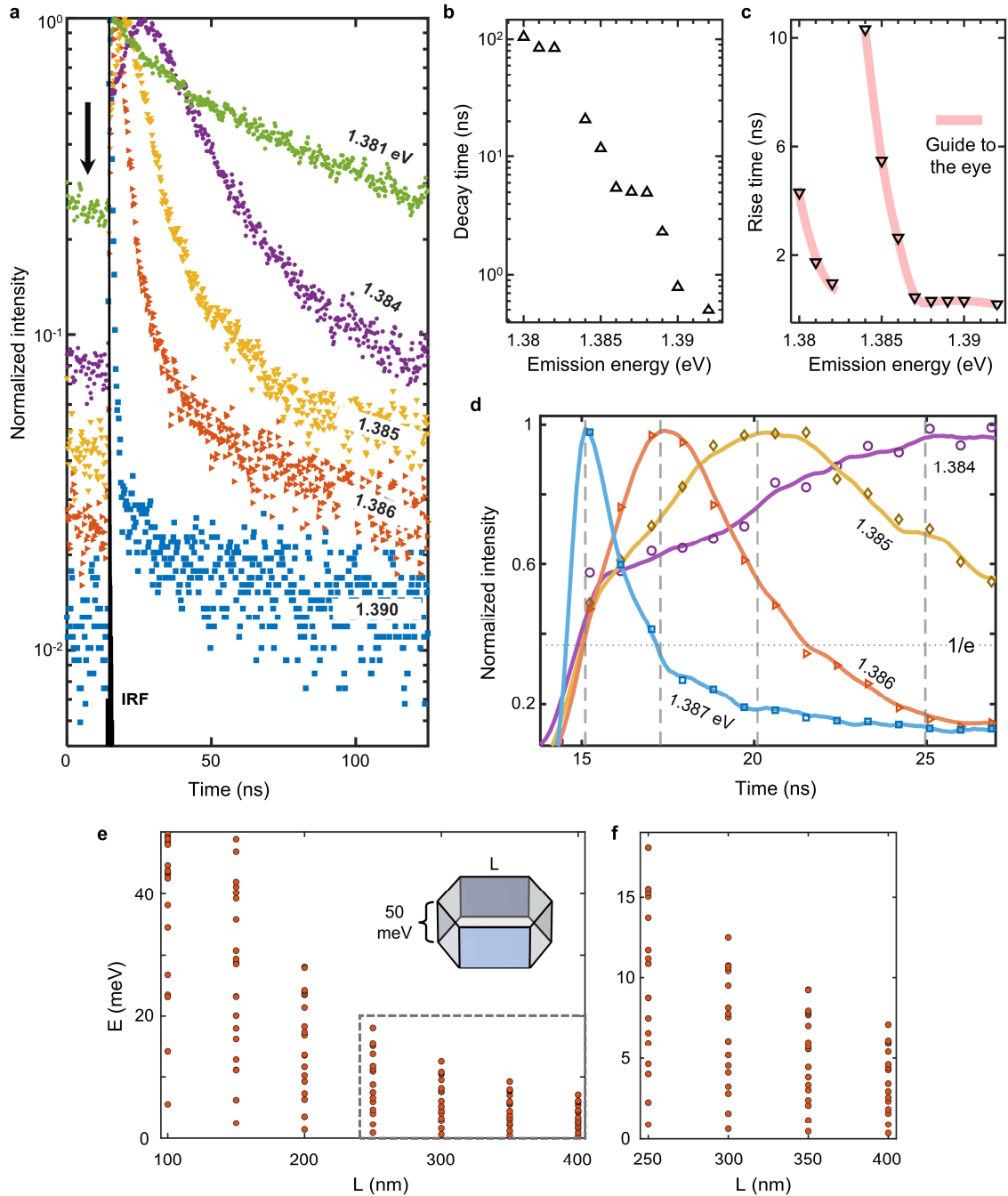

**Fig. 2 | Time-resolved emission spectra at 8 MHz excitation.** (a) Time Resolved photoluminescence (TRPL) from different IELX emission energies (as labelled) with 8 MHz repetition rate excitation is plotted in semi-log scale. Black arrow indicates the increasing residual population for long lived IELX states. Dark black filled curve near 15 ns is the instrument response function (IRF). (b) Decay time

(semi-log scale) for different emission energies. (c) Rise time (when the intensity reaches maximum with respect to the IRF) for different emission energies. (d) Close-up of the TRPL from four consecutive emission energies (as labelled) shown in linear scale. The dashed black lines represent the rise time, and dotted line shows the 1/e value on the intensity scale. (e) Calculated energy levels of an exciton in a hexagonal well of different side lengths, for a finite well of height 50 meV, and (f) zoomed in, around 300nm side length. Inset in Fig. (e) is schematic of the potential well.

To understand the origin of multiple emission peaks under 8 MHz excitation, we perform TRPL at different emission energies (Fig. 2a and Supplementary Section 6). Emission from the sample is filtered through the spectrometer and subsequently detected by a single photon avalanche diode (SPAD) to measure TRPL. This setup allows us to measure TRPL with ~ 0.7 meV energy resolution. As we scan from higher to lower emission energies, we observe drastically varying dynamics (Fig. 2a). The 1.390 eV emission exhibits extremely fast rise and decay time (sub-nanoseconds (ns)), while 1.381 eV emission takes ~ 2 ns to reach maximum intensity and decays with ~ 100 ns time constant. Remarkably, within a range of ~ 10 meV, the decay time has decreased monotonically by two orders of magnitude (Fig. 2b). Thus, the drastically different dynamics confirms the finely spaced PL peaks are different levels under the emission envelope demonstrating quantum confinement (Fig. 1e). Further, this immense lifetime variation can be used in quantum sensing devices to sense the dielectric environment.

For 80 MHz repetition rate, the laser pulses excite the sample every 12.5 ns, much faster than the 100 ns decay time of the lower energy IELX. This fast excitation rate leads to saturation and effective suppression of emission from these states, as evidenced by the build-up of a residual population, indicated by the black arrow in Fig. 2a. Thus, in time-integrated PL (Fig 1e), a lower (higher) PL from lower energy states for 80 MHz (8 MHz) excitation is observed. Notably, rise time also shows a significant non-monotonic variation with emission energy (Fig. 2c), which we discuss later.

Remarkably, at closer inspection of the first 15 ns window (Fig. 2d), particularly in emissions range 1.384-1.387 eV, we find the decay of each state is closely followed by the rise of the closest lower-energy state. For example, the 1.387 eV emission intensity crossed the 1/e value close to when 1.386 eV emission reaches its maximum intensity. This indicates the sequential filling of the lower state by the higher state. A similar trend is present for other emission energies as well, but with a slightly smaller rise time, likely due to presence of multiple states above the concerned emission states. Such dynamics are called cascade transitions, and have been previously observed in quantum wells[25] and moiré systems[26]. In the presence of multiple energy states in a single well, higher energy states are filled first, followed by a decay either to the ground state (emitting a photon), or to lower energy states, shown in Fig. 4(a). Such dynamical correlation between the energy states is considered to be direct proof that the energy states are part of a single potential well, instead of different potential wells[26]. Hence, this observation confirms the presence of multiple dynamically correlated energy states in large, reconstructed $H_h^h$ domains. We note that a mobility-edge based argument can be employed in case of quantum dots, but for a continuous distribution of states (inhomogeneous distribution)[27].

To further support our observation, we compare the observed fine energy spacing with the theoretical expectation for a quantum particle of mass *m* in a hexagonal well of side *L*, a probable shape of the mesoscopic domains. The quantum mechanical problem of a particle in an infinite potential well of a regular polygon shape is exactly solvable (for a square and an equilateral triangular case). For example, in the case of a triangular well, one obtains the energy spectrum $E_{p,q} = \frac{8\pi^2 \hbar^2}{3mL^2}(p^2 + 3q^2 + 3pq)$, where *q* are integer multiples of 1/3 and *p* are positive integers[28]. Here, we numerically solve this problem for a hexagonal potential well with different *L*. We use the effective mass of the exciton *m = 1.15 $m_e$*, which is obtained as the sum of the effective electron mass of MoSe$_2$ and the effective hole mass of WSe$_2$. The values of the respective electron and hole effective mass are taken from first-principles data by Förg *et. al.*[29]. From these calculations, we find that for a well of side length 300 nm and potential height of 50 meV (Fig. 2e and 2f), the obtained average energy spacing is about 1 meV, consistent with experimental observation. Energy levels for infinite potential wells (Supplementary Section 9) obtain a similar energy spectrum. This similarity is expected as potential height of 50 meV is well above the lowest 20 energy levels, which are spaced by about 1-1.5 meV. However, for smaller $L$, when the energy spacing increases such that higher energy levels approach the well height of 50 meV, the energy levels of the finite well begin to differ significantly from those of the infinite well.

Interestingly, the 1.381 eV emission reaches maximum intensity significantly earlier than the 1.384 eV state (Fig. 2c). This suggests that these two emission states are not directly connected to each other. Also, the 1.380-1.382 eV emission shows a consistent increase in rise time with decreasing emission energy, similar to the trend in 1.383-1.387 eV range. From this understanding, we speculate that the two sets of emission, 1.380-1.382 eV, and 1.383-1.387 eV arise from two independent potential wells, with different domain sizes.

## Quantum Siphoning: competing radiative and non-radiative processes

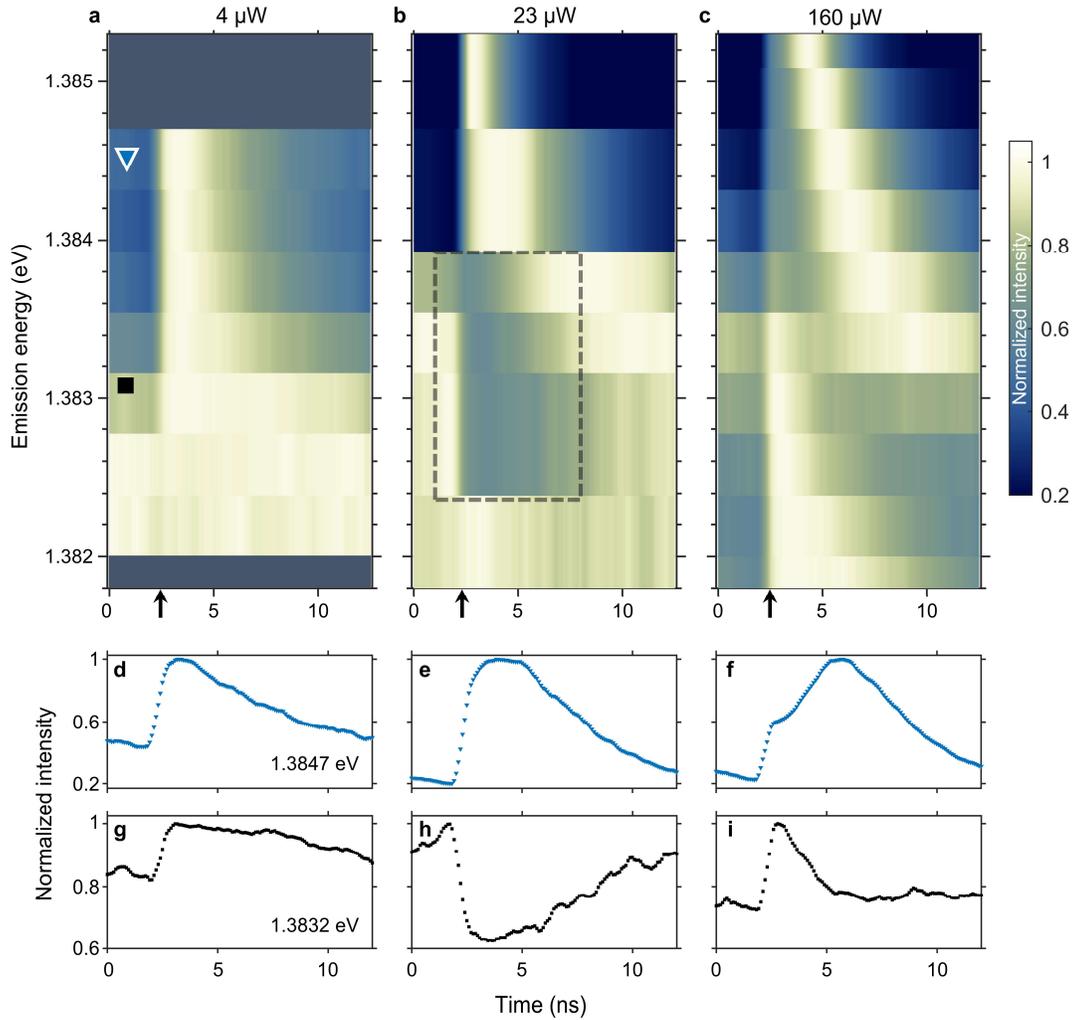

**Fig. 3 | Power-dependent time-resolved emission spectra at 80 MHz excitation.** (a) to (c) Emission energy resolved TRPL maps for excitation with 4, 23, and 160 µW average power for 80 MHz repetition rate excitation. Dashed rectangle indicates the region where we observe the TRPL dip. Black arrow indicates the excitation laser timing. The triangle and square marker in Fig. 3(a) indicate 1.3847 and 1.3832 eV emission energies, respectively. (d) to (f) [(g) to (i)] Linecuts from the maps at 1.3847 [1.3832] eV emission energy for 4, 23, and 160 µW average power, respectively.

Next, we perform emission-energy resolved TRPL measurements with 80 MHz repetition rate laser. Dynamics for the different emission energies are shown in Fig. 3(a) to (c) for average excitation powers of 4, 23, and 160 µW, respectively. Time-integrated PL under these conditions is shown in Supplementary Section 8 (Fig. S9a). At low power (4 µW, Fig. 3a), the TRPL map shows decreasing residual population from low to high-energy emission, reflecting shorter decay time at higher energies. For instance, the lower-energy emission at 1.3832 eV showed a residual population of about 80%, significantly higher than the 40% observed for the higher-energy emission at 1.3847 eV, due to its longer lifetime (Fig. 3d and 3g).

At medium power (23 µW, Fig. 3b), dynamics becomes completely different as we go from low to high emission energies. For the 1.3847 eV emission (Fig. 3e), intensity increases at laser excitation and then decays, similar to behaviour seen for 8 MHz excitation (Fig. 2). Increase in non-radiative recombination led to faster recombination dynamics and less residual population than at low power. In contrast, for 1.3832 eV emission (Fig. 3h), the intensity decreases ~ 40% immediately after excitation, followed by a slow rise to the maximum TRPL intensity. The dip is more prominent where the residual populations are larger at lower power. Further power-dependent data is shown in the Supplementary Section 8. This dip indicates a nonlinear dynamical feature, observed earlier in a $MoS_2/WSe_2$ moiré system with ~ 20 meV energy level spacing[26]. Remarkably, we observe this dramatic feature in extremely finely spaced levels (~ 1 meV) in the reconstructed system. This observation further solidifies the presence of multiple closely spaced energy levels within a single potential well. Tan et al. reasoned that upon laser excitation, the residual excited population will be re-excited to the highest energy state, decreasing the PL and leading to the TRPL dip[26]. However, in this scenario, the TRPL dip should be present for any arbitrary residual population, which is not observed in our experiments. Moreover, excited state absorption can potentially enable photon upconversion, but we have not observed any new emission in the spectral range (1.77-3.1 eV) above our excitation energy[30]. The observation of similar features in two drastically different localization scenarios (moiré vs reconstructed domains) suggests that the observed nonlinear phenomenon is quite general in nature, with interesting consequences on many-body physics and correlations[31].

The power-dependent TRPL measurements suggest that a large residual population alone is not enough to produce the PL dip. Theoretical calculations for the $MoSe_2/WSe_2$ system predict direct K-K transitions, as well as indirect K-Q and K-Γ valley transitions[32–34]. Experimental studies show that, at high excitation densities, excitons get scattered from K valley to Q and Γ valleys[35–37]. Here, we also note that the TRPL dip is not observed for lower-energy long-lived states (~1.38 eV), even with large residual population at lower powers (Supplementary Section 8). As evidenced by lower energy and longer lifetimes, these states are much more trapped and hence possess lower oscillator strength, leading to lower interaction with the excitation laser[38].

We suggest that the PL dip arises from exciton population scattering from the radiative states to non-radiative intervalley states, when sufficient residual population, excitation density, and large oscillator strength are present (Fig. 4a). We term this phenomenon as "quantum siphoning" effect, similar to a Pythagorean cup "siphoning" its contents away when a threshold amount is crossed (Fig. 4b). Further, quantum siphoning can be used to engineer novel excitonic dynamics, for example using strain engineering to control valley scattering, or by creating strain-modulated mesoscopic domains[39]. On the other hand, for 8 MHz excitation, we observed cascade transitions within ~ 1 meV separated energy states, with accompanying population transfer to dark states[26]. With increasing excitation density, these dark and bright states get saturated due to finite density of states, resulting in an increased residual population (see Fig. 2a).

At high excitation power (160 μW, Fig. 3c), the TRPL dip is no longer observed in any measured emission region (Supplementary Section 8). Instead, a clear signature of the cascade transition of two distinct emission sets (as observed for 8 MHz excitation) is visible in the TRPL map (Fig. S8e). The 1.3847 eV emission (Fig. 3f) shows a slow rise followed by a decay. Interestingly, 1.3832 eV emission (Fig. 3i) exhibits a bi-exponential nature, with a faster and slower decay component (similar to low excitation power). This complex power dependence can be explained by two processes: At higher IELX densities, increased non-radiative recombination leads to an initial fast decay[40,41], followed by a slower decay. Secondly, the saturation of Q and Γ valleys at high powers can reduce scattering to these valleys (from K valley). Hence, a competition between the reduction of residual population and scattering probability results in the dynamics at high powers.

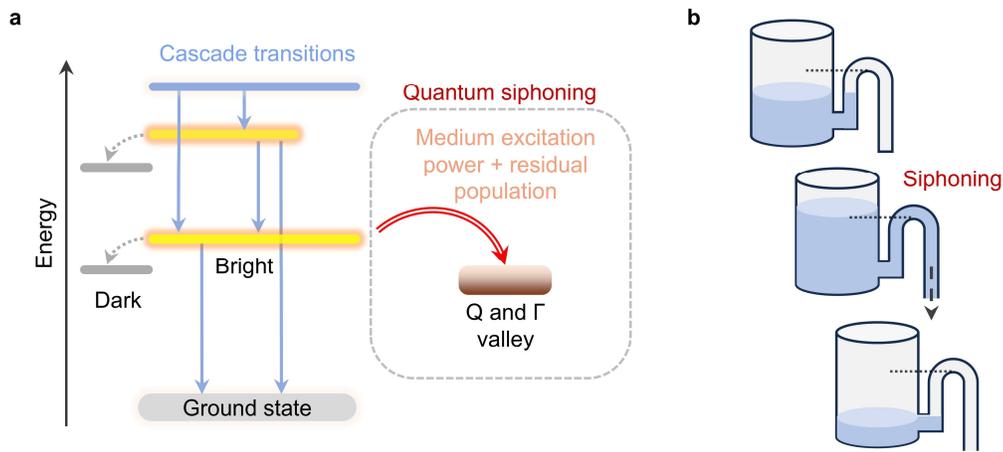

**Figure 4 | Schematic representation of cascade transitions and quantum siphoning.** (a) The schematic depicts a simplified picture of different bright and dark energy levels. At low-repetition rate excitations, cascade transitions occur, as shown by the blue arrows. Black arrows show part-channelling of excitations into dark states. Under medium excitation power and residual population, upon laser excitation, a large amount of excitons scatter from bright K valley states to non-radiative Q and Γ valleys, resulting in a transient reduction of PL, indicated as quantum siphoning. (b) A Pythagorean cup: From top to bottom figure, when the liquid level reaches the threshold (dotted line), the liquid siphons out of the tube, making the cup empty (dashed arrow).

**Conclusion:**

In summary, we demonstrate multiple closely-spaced energy levels in reconstructed mesoscopic domains. Firstly, this demonstrates the impact of quantum confinement in flat potential landscapes. Remarkably, emission energy resolved TRPL reveals two orders of magnitude lifetime variation within a narrow ~ 10 meV range, potentially explaining the longstanding discrepancies in reported emission dynamics. Cascade transitions between closely spaced states (~1 meV) further establish their origin from a single potential well, with distinct emission from additional potential wells. Next, TRPL at a higher excitation rate results in a large residual population due to the long lifetime of decay channels and saturation. Under this condition, TRPL uncovers a transient reduction of PL driven by scattering of bright K valley excitations into non-radiative Q and Γ valleys, followed by gradual recovery, a phenomenon we introduce as quantum siphoning.

The drastic variation of lifetime in a narrow emission window shows the dramatic influence of quantum confinement, thus opening this system for quantum sensing applications. Our approach—leveraging repetition-rate-dependent excitation to reveal hidden emission states and novel intervalley scattering dynamics—provides a general platform for uncovering subtle excitonic phenomena in reconstructed 2D heterostructures, and can be extended to strain- and defect-engineered systems. Our identification of the quantum siphoning process underscores the need for theoretical models addressing scattering-dominated processes, and can enable control of exciton dynamics via strain engineering.

40. Sun, D. *et al.* Observation of Rapid Exciton–Exciton Annihilation in Monolayer Molybdenum Disulfide. *Nano Lett.* **14**, 5625–5629 (2014).

41. Wu, J.-M. *et al.* Exciton luminescence and many-body effect of monolayer WS2 at room temperature. *Chinese Phys. B* **31**, 057803 (2022).

**Methods:**

**Sample Preparation**: MoSe$_2$ and WSe$_2$ bulk crystals were sourced from 2D Semiconductors, while hexagonal boron nitride (hBN) was obtained from NIMS, Japan. Bulk TMD crystals were mechanically exfoliated using Nitto blue tape. To transfer flakes, a small PDMS sheet (Gel Pak PF-40X40-0065-X4) was applied to the tape and brought into contact with a SiO$_2$/Si substrate. hBN was exfoliated using established methods[42]. TMD monolayers were identified using the RAW imaging-based layer-identification method[43]. Heterostructures were assembled using a PDMS-PC stamp technique, with each layer picked up at 110°C and the final drop performed at 170°C[17]. Following assembly, samples were cleaned by immersion in chloroform and isopropyl alcohol for five minutes each, followed by drying with nitrogen. To promote atomic reconstruction and improve interface quality, samples were annealed at 250°C for three hours in a glove box under inert atmosphere[44].

**Optical experiments**: Room temperature integrated PL map is measured using a Picoquant Micro Time 200 setup. The cryogenic PL is measured using a custom-built free-space optical setup. For changing the repetition rate of the excitation laser, we have used the output of a pulsed laser (with 100 fs pulse width and 80 MHz repetition rate, Ti-Sapphire Laser, Mai Tai HP) and passed it through a pulse picker (Spectra-Physics model 3980). Using the pulse picker, the repetition rate was varied from 1 to 80 MHz. Focusing of excitation source and emission collection are done using a 50x Mitutoyo long-working distance objective, with 0.42 numerical aperture, placed outside the cryostat. The emission is then sent to the Andor Kymera 328i spectrometer, dispersed by a 300 or 900 lines/mm grating, and detected using an iDus 416 CCD. For TRPL measurements, the emission is passed through the spectrometer (used as a monochromator) and focused on a single-photon avalanche diode (Micro Photon Devices), equipped with a PicoHarp 300 (PicoQuant) timing module. For second harmonic generation, we used the 820 nm pulsed laser with 80 MHz repetition rate and focused it onto the sample using a Mitutoyo 100X long working distance objective (NA = 0.5), producing a 2 μm diameter laser spot.

**Calculation of energy levels**: To calculate the energy levels of a quantum particle in a regular polygon potential well of side L, we consider a larger square space of side length 4L and divide this space into a finely spaced square grid (201 × 201 grid points), such that each grid point is positioned at

$$R_{ij} = \left(\frac{i}{4L}, \frac{j}{4L}\right)d, \quad i,j \in [0,1,\ldots,N), \quad (1)$$

where $d$ is the distance between nearest neighbor grid points. We use the finite difference method to discretize the Laplacian as

$$\nabla^2 \psi_{i,j} \approx \frac{\psi_{i+1,j} + \psi_{i-1,j} + \psi_{i,j+1} + \psi_{i,j-1} - 4\psi_{i,j}}{d^2}. \quad (2)$$

At the square boundary, we use finite-size boundary condition so that in the absence of any other potential, it behaves as an infinite square well. Next, we consider the finite well potential by considering the hexagon of side $L$ centered at the origin and include the local potential for points $R_{i,j}$ outside the hexagon. We numerically solve the eigenvalue equation

$$\sum_{i,j}\left(V_{i,j} + \frac{2\hbar^2}{md^2}\right)\psi_{i,j} - \frac{\hbar^2}{2md^2}(\psi_{i+1,j} + \psi_{i-1,j} + \psi_{i,j+1} + \psi_{i,j-1}), = E\sum_{i,j}\psi_{i,j} \quad (3)$$

to obtain the energy spectrum. We have checked the numerical accuracy by choosing a very deep triangular well potential and verifying the numerically obtained energy levels against the known exact results of the infinite triangular well potential.

**Data Availability:**

All data created or analysed during this study are included in this paper and in the Supplementary Information. Additional supporting data are available from the corresponding author upon reasonable request.

**Acknowledgements:**

AS and MM thank Manish Jain for discussions and insights, and A.K. Sood for the pulse picker system. MM thanks the Prime Minister's Research Fellowship (PMRF) for the Ph.D. fellowship. AS acknowledges funding from Department of Science and Technology Nanomission CONCEPT grant (NM/TUE/QM-10/2019), Anusandhan National Research Foundation grant (CRG-2022-003334), and MoE-STARS (STARS-2/2023-0265). GC acknowledges support from EPSRC ERC underwrite grant EP/X025829/1.

**Author Information:**

Corresponding Author: *Akshay Singh, aksy@iisc.ac.in

**Author Contributions:**

MM and AS developed the experimental framework. MM performed sample preparation, and optical measurements. MM and AS discussed and analysed the data. KW and TT synthesized the hBN crystals. GC carried out theoretical calculations. MM, GC, and AS discussed and prepared the manuscript.

**Competing interests:**

The authors declare no competing interests.

**Additional Information:**

**Supplementary Information** is available for this paper.

**Correspondence and requests for materials** should be addressed to Akshay Singh.

Supplementary Information

Quantum siphoning of finely spaced interlayer excitons in reconstructed MoSe$_2$/WSe$_2$ heterostructures

Mainak Mondal[1], Kenji Watanabe[2], Takashi Taniguchi[3], Gaurav Chaudhary[4], Akshay Singh[1*]

[1]Department of Physics, Indian Institute of Science, Bengaluru, India

[2]Research Center for Functional Materials, National Institute for Materials Science, Japan

[3]International Center for Materials Nanoarchitectonics, National Institute for Materials Science, Japan

[4]TCM Group, Cavendish Laboratory, University of Cambridge, Cambridge, United Kingdom

Corresponding Author: *aksy@iisc.ac.in

Sections:

1. Room temperature photoluminescence (PL) map
2. Second Harmonic Generation
3. Temperature dependent PL
4. Fitting of CW laser excitation PL
5. Power dependent PL for CW laser excitation
6. Time Resolved PL (TRPL) for different emission energies at 8 MHz
7. Repetition rate dependent PL
8. Emission resolved PL and TRPL data for different excitation power at 80 MHz
9. Energy level calculations

**Section 1: Room temperature photoluminescence (PL) map**

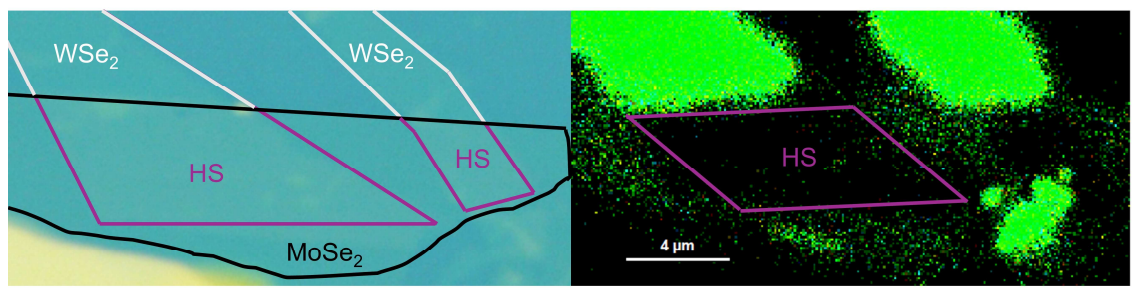

Figure S1: Left side: optical image of the device is shown with marked WSe$_2$, MoSe$_2$ and HS region. Right side: integrated PL map at room temperature shows highly quenched PL from the HS region, indicating good interface quality[1].

**Section 2: Second Harmonic Generation**

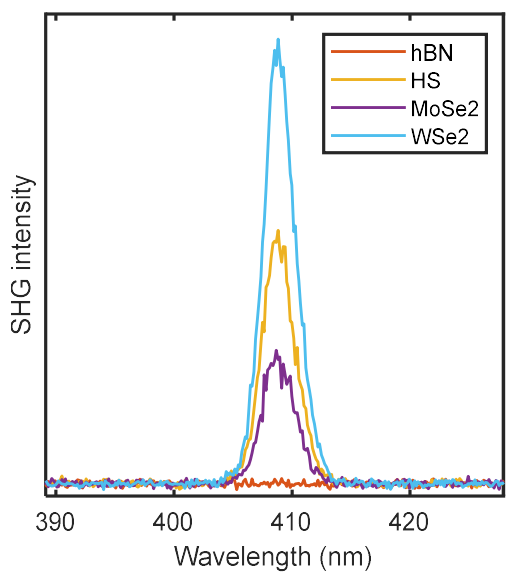

Figure S2: Second harmonic generation (SHG) measurements on the MoSe$_2$, WSe$_2$ and on the heterostructure (HS) part (labelled region on the optical image), shows a decrease in SHG signal on the HS, confirming the H-type stacking in the device, matching with previous reports[2,3]. Measurements are performed using 820 nm excitation with 100 fs pulse width and 80 MHz repetition rate, without changing the analyser polarization.

**Section 3: Temperature dependent PL**

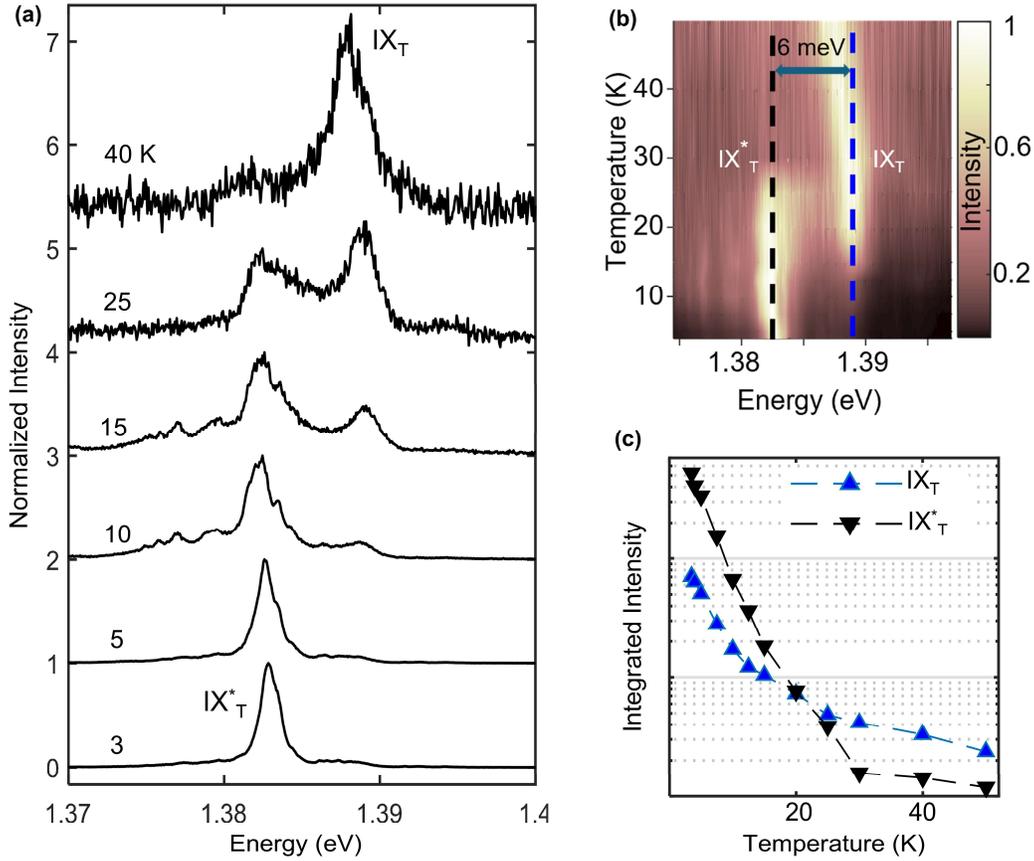

Figure S3: (a) Normalized PL for different temperatures (as labelled). (b) Normalized PL map with increasing temperature shows disappearance of lower energy peak ($IX_T^*$) and a sustained higher energy peak ($IX_T$) with energy difference ~ 6 meV. (c) Integrated intensity from both peaks plotted with temperature.

With increasing temperature, the emission intensity shows two very prominent features. Firstly, as temperature goes above 20 K the peak at ~ 1.383 eV almost disappears, but emission from ~ 1.389 eV (at 25 K) sustains which is evident from the normalized PL shown in Fig S3a and S3b. The energy difference between these two peaks is ~ 6 meV, matches very well with the reported energy difference for interlayer triplet exciton and triplet trion[4–6]. The dominant trion population at low temperatures is due to unintentional doping present in the system. Secondly, the emission intensity from the trion peak decreases very promptly (~ 100 time for an increase of 20 K) suggesting very shallow potential wells, also previously reported for interlayer trions (Fig. S3c)[7,8]. In combination, these measurements confirm that the observed emission originates from interlayer triplet trion emission in the reconstructed domains in a MoSe$_2$/WSe$_2$ HS with shallow potential depths.

**Section 4: Fitting of CW laser excitation PL.**

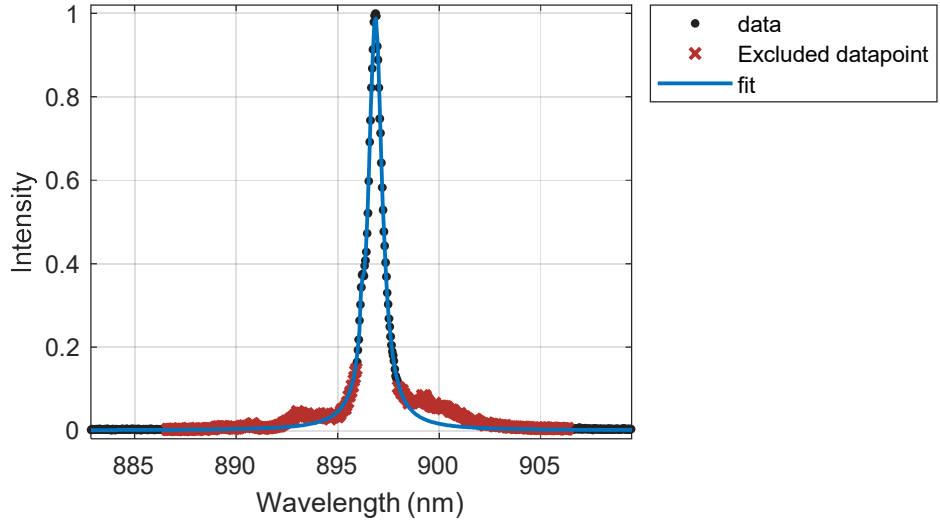

Figure S4: PL at 4K temperature with 478 nm continuous wave laser at 0.64 µW power.

The PL is fitted with two Lorentzian functions, $y = \frac{a1*w1^2}{(x-c1)^2 + w1^2} + \frac{a2*w2^2}{(x-c2)^2 + w2^2}$

The fitting results are

| a1 | w1 (nm) | c1 (nm) | a2 | w2 (nm) | c2 (nm) |
|---|---|---|---|---|---|
| 0.9926 | 0.4021 | 896.9 | 0.1081 | 0.0786 | 896.2 |

Here the spectrum is mostly dominated by a single Lorentzian peak as can be seen from the fitted parameters.

**Section 5: Power-dependent PL for CW laser excitation.**

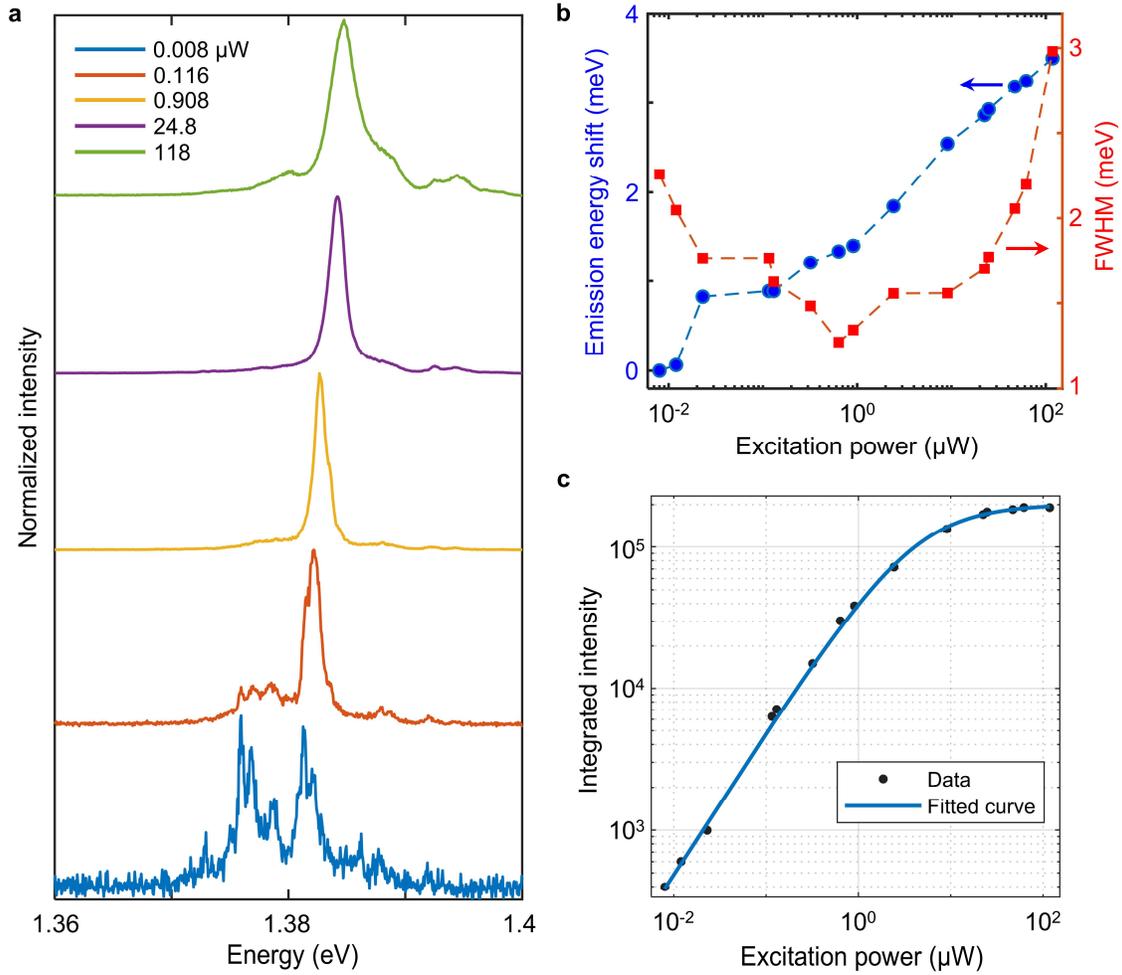

Figure S5: (a) The variation of PL with increasing power (labelled in the figure) is shown at 4K temperature. (b) The shift of central energy peak and variation of full width half maximum (FWHM) with power. (c) Saturation of integrated intensity with increasing power.

PL collected at 4 K (478 nm continuous wave laser) with different excitation powers is shown in Figure S5a. The emission shows a FWHM of ~ 1.5 meV with a nearly Lorentzian line shape (fitting shown in SI section 4), a characteristic emission from reconstructed $H_h^h$ domains[4,6,9,10], which is significantly lower by 20 meV compared to non-reconstructed samples[11]. Also, increasing the excitation power (up to 4 order) causes a total ~ 3 meV blue shift of emission energy and a slight increase in emission spectral width (Figure S5b), showing great similarities with the reported emission from reconstructed $H_h^h$ domains[5]. The integrated intensity is taken as area under the curve from 1.380 to 1.390 eV. The integrated intensity is fitted with, $y = \frac{k*power}{power + p_{sat}}$, here $power$ is excitation power and $p_{sat}$ is 4.132 µW, which represents the saturation power.

**Section 6: Time Resolved PL (TRPL) for different emission energies at 8 MHz**

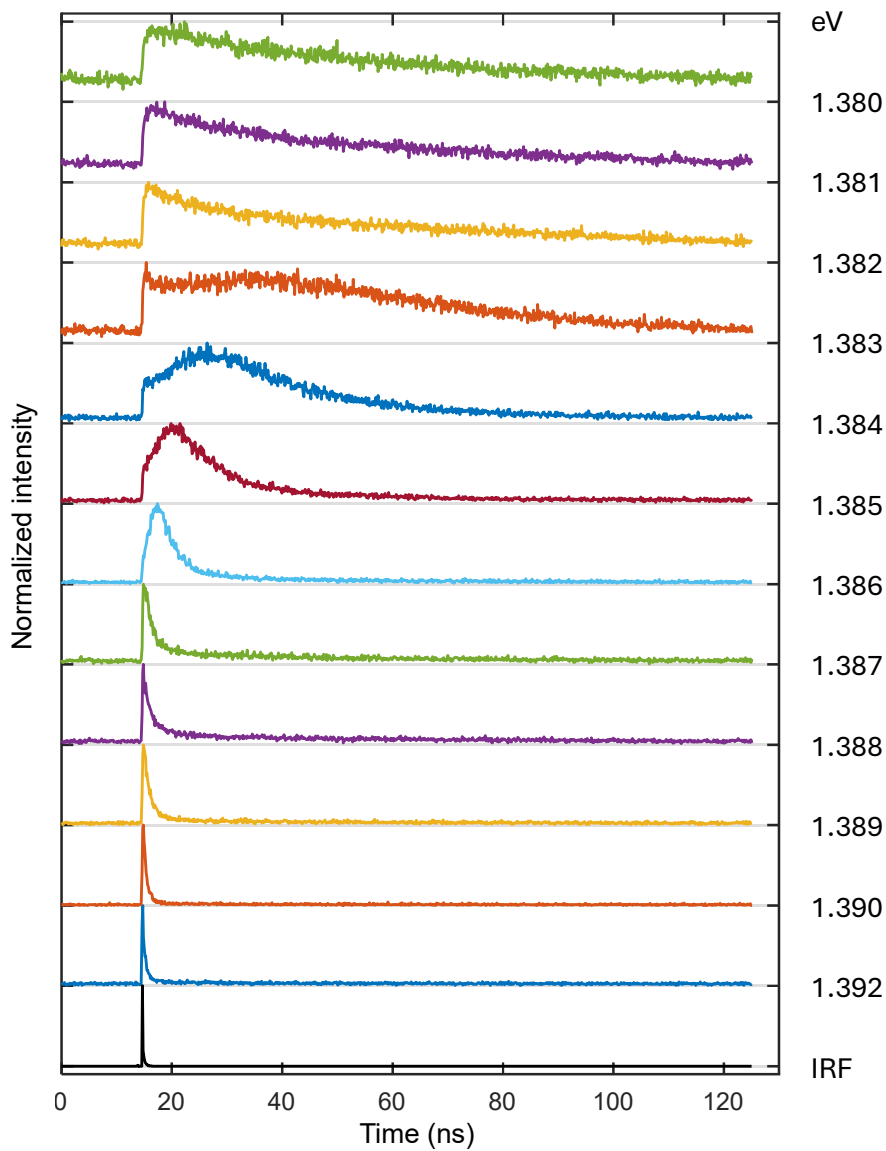

Figure S6: Instrument response function (IRF) was measured with 4 ps resolution, a 32-element moving average is taken to plot it with the data which was taken with 128 ps resolution. The emission energy is labelled on the side.

**Section 7: Repetition rate dependent PL**

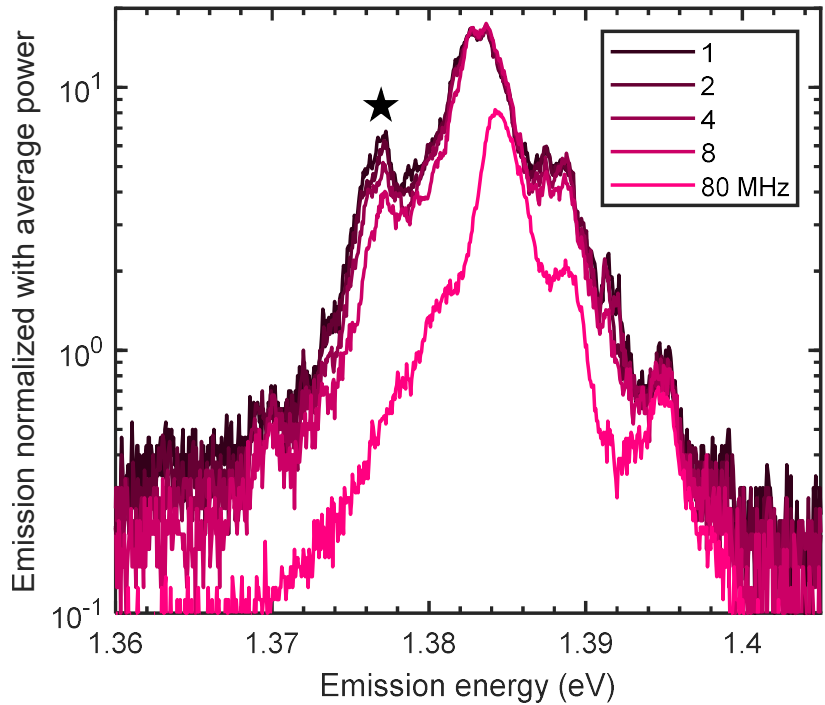

Figure S7: Emission for different excitation repetition rates with the same pulse peak power is divided with average power to normalize it and plotted in log scale.

Emission intensity has decreased more for lower energy than the higher energy region for 80 MHz, due to higher lifetime and resulting saturation. The peak at ~ 1.376 eV (marked with a star) shows systematic decrease in emission intensity with increasing repetition rate.

**Section 8: Emission resolved PL and TRPL data for different excitation power at 80 MHz**

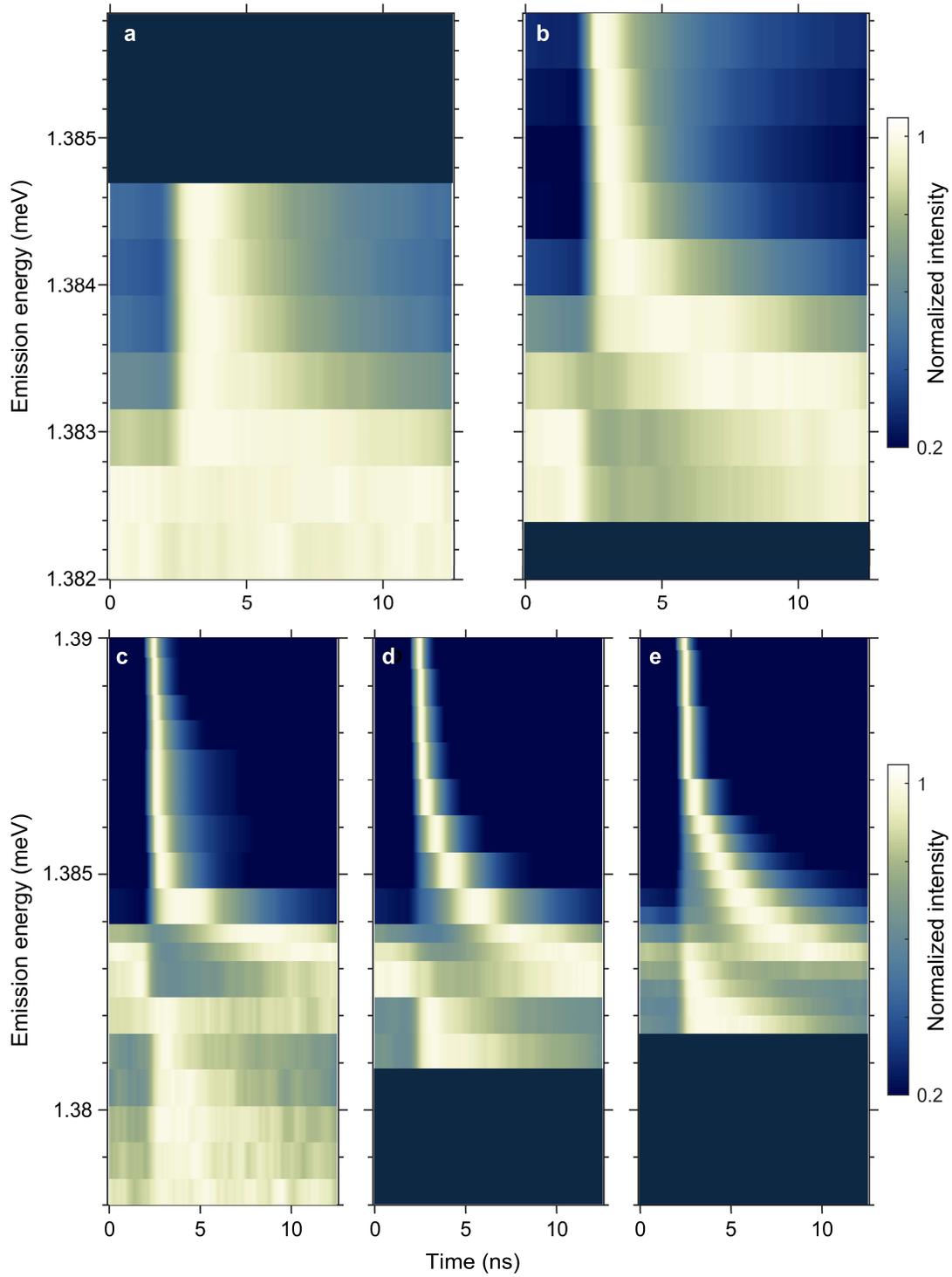

Figure S8: Complete emission energy resolved TRPL maps are shown for (a) 4, (b) 11, (c) 23, (d) 120, (d) 160 µW excitation powers with 725 nm laser at 80 MHz repetition rate.

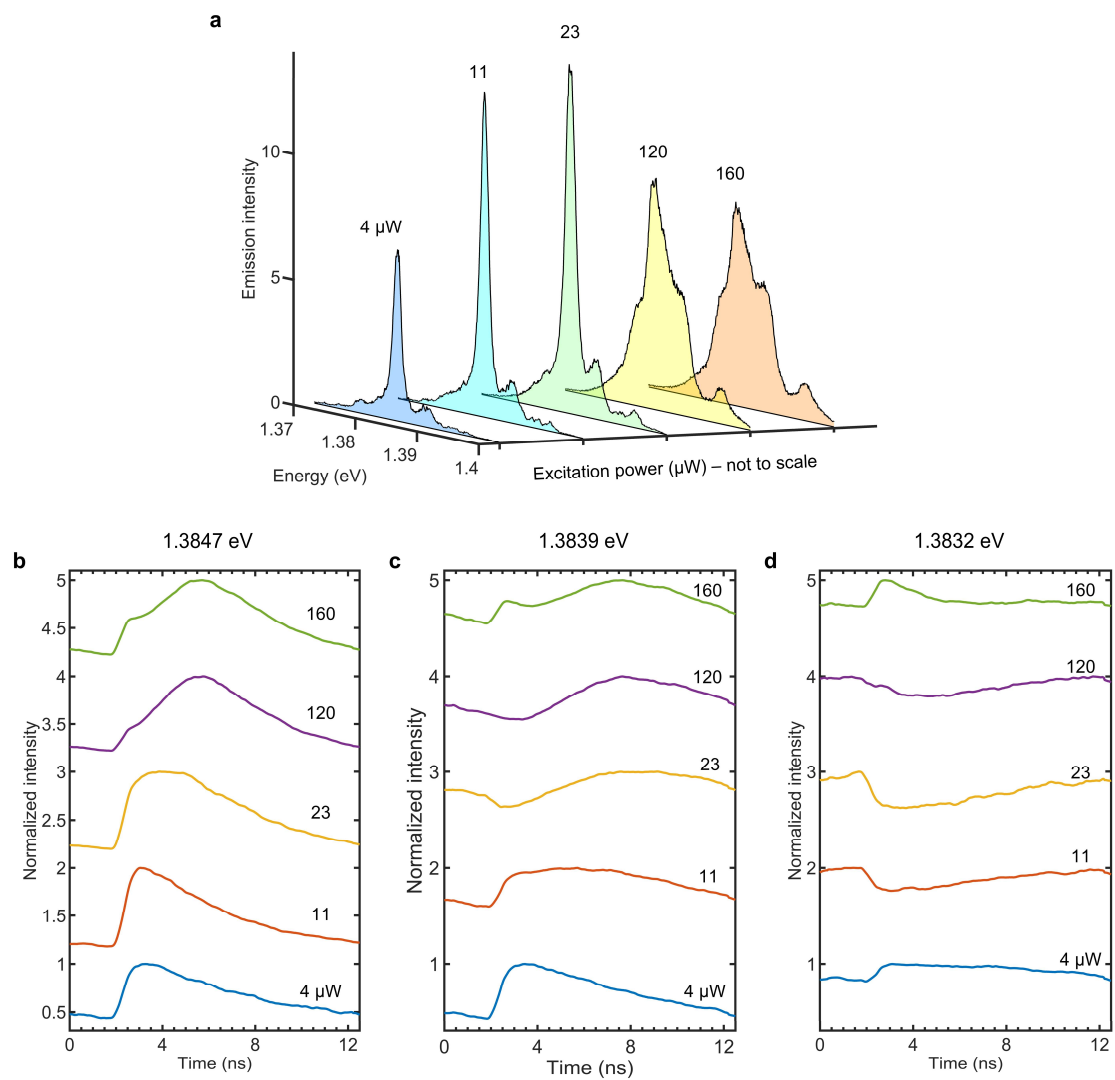

Figure S9: (a) PL emission for different excitation powers. (b), (c) and (d) stacked linecuts from the emission resolved TRPL maps shown in Fig. 3 in main text and Figure S8.

**Section 9: Energy level calculations**

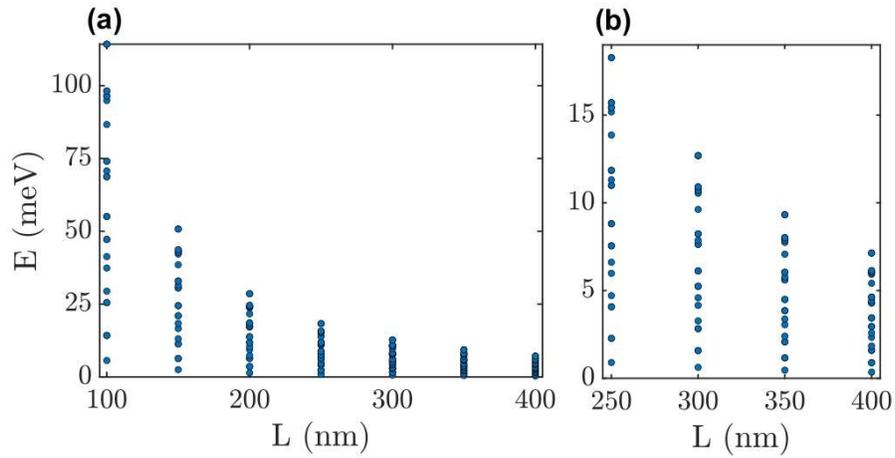

Figure S10: (a) Energy levels of a quantum particle (with mass of exciton) in a hexagonal well of different side lengths for an infinite well. (b) zoomed in around 300 nm side length.